\documentclass[reprint,superscriptaddress, amsmath,amssymb, floatfix]{revtex4-2}
\usepackage{graphicx}
\usepackage{amsmath}
\usepackage{siunitx}
\usepackage{caption}
\usepackage{ragged2e} 
\usepackage{bm}
\usepackage{textcomp, gensymb}
\usepackage{hyperref}
\hypersetup{colorlinks = true,linkcolor=blue,
     filecolor=blue,
     urlcolor=blue,
     citecolor = blue, pdfauthor=author}
\begin{document}
\captionsetup[figure]{labelfont={default},labelformat={default},labelsep=period,name={Fig.}}
\renewcommand{\figureautorefname}{Fig.}

\title{Space-Time Elastic Metamaterials for Zero-Frequency and Zero-Wavenumber Bandgaps}

\author{Brahim Lemkalli}
\email{lemkallibrahim@gmail.com}
\affiliation{Université de Franche-Comt\'{e}, Institut FEMTO-ST, CNRS UMR 6174, Besan\c{c}on, 25000, France}
\author{Alaa Ali}
\affiliation{Université de Franche-Comt\'{e}, Institut FEMTO-ST, CNRS UMR 6174, Besan\c{c}on, 25000, France}
\author{Qingxiang Ji}
\affiliation{Université de Franche-Comt\'{e}, Institut FEMTO-ST, CNRS UMR 6174, Besan\c{c}on, 25000, France}
\author{Julio Andrés Iglesias Martínez}
\affiliation{Université de Lorraine, Institut Jean Lamour, CNRS UMR 7198, Nancy 54000, France}
\author{Younes Achaoui}
\affiliation{OPTIMEE Laboratory, Department of Physics, Moulay Ismail University, B.P. 11201, Zitoune, Meknes, Morocco}
\author{Sebastien Guenneau}
\affiliation{UMI 2004 Abraham de Moivre-CNRS, Imperial College London, SW7~2AZ, London, UK}
\author{Richard Craster}
\affiliation{UMI 2004 Abraham de Moivre-CNRS, Imperial College London, SW7~2AZ, London, UK}
\author{Muamer Kadic}
\affiliation{Université de Franche-Comt\'{e}, Institut FEMTO-ST, CNRS UMR 6174, Besan\c{c}on, 25000, France}

\begin{abstract}
We create wave-matter space-time metamaterials using optical trapping forces to manipulate mass-spring chains and create zero-frequency and zero-wavenumber band gaps: the bosonic nature of phonons, and hence this elastodynamic setting, traditionally prohibits either zero-frequency or zero-wavenumber band gaps. Here, we generate zero-frequency gaps using optomechanical interactions within a 3D mass-spring chain by applying an optical trapping force to hold or manipulate a mass in a contactless manner independent of its elastodynamic excitations. Through careful modification of the geometrical parameters in the trapped monoatomic mass-spring chain, we demonstrate the existence of a zero-frequency gap generated by the optical forces on the masses. The precise control we have over the system allows us to drive another set of masses and springs out of phase with its traveling wave thereby creating a zero-wavenumber band gap. 
\end{abstract}

\maketitle


For centuries, researchers have focused on how acoustic and electromagnetic waves interact with matter to better understand material  properties and govern wave propagation; these investigations are critical for a variety of applications, ranging from understanding the origins of the universe to developing new devices and technology. The field of photon-matter interaction, which studies how electromagnetic waves interact with matter, has grown beyond its initial boundaries and now involves the application of mechanical stress on matter via the energy, linear momentum, and angular momentum of light. Historically, Johannes Kepler was the first to identify this phenomenon in 1619 by observing the deflection of comet tails by sunlight, thereby establishing that light can exert mechanical forces on matter \cite{Jackson2003}.
Building on this concept, Arthur Ashkin introduced the application of optical forces in 1970, leading to the development of optical tweezers \cite{Ashkin197024, Ashkin197025, ashkin86}.
The use of light for trapping microparticles has since led to groundbreaking discoveries and diverse applications in fields such as chemistry \cite{Sullivan2020, Shen2022, Zaltron2020}, nanotechnology \cite{Maragò2013, Melzer2021,Mcleod2009ArraybasedON, Mcleod2008SubwavelengthDN}, quantum science \cite{Kaufman2021}, plasmonics \cite{Zhang2021}, thermodynamics  \cite{Gieseler2018}, and microrobotics \cite{Gerena2019,Zhang20208}.\\
In parallel, the field has also transformed into a dynamic exploration of phonon vibrations within phononic crystals and metamaterials \cite{craster2023mechanical,kadic20193d,chen2020isotropic}. This research aims to generate stop band gaps, which are generally essential for a variety of applications, such as shielding \cite{miniaci2016large, luo2023surface}, sensing \cite{alrowaili2023heavy, almawgani2023periodic, lucklum2012two}, and filtering \cite{pennec2004tunable, chen2017acoustic}.  
Inspired by photonic crystals, phononic structures control the propagation of phonons using an array of periodic scatterers and create high-frequency band gaps via Bragg scattering, with the periodic constant matching the wavelength of the incident wave \cite{kushwaha1994band, de1998ultrasonic}. Notably, creating low-frequency band gaps with periodic arrays of scatterers was challenging. However, in 2000, the concept of local resonance emerged, revolutionizing band gap creation by introducing a mechanism where propagating modes interact with locally resonant modes, resulting in low-frequency local band gaps \cite{liu2000locally}. Following this, metamaterials with sub-wavelength band gaps arose, owing to the aforementioned mechanisms. Since then, substantial optimization has permitted the development of low-frequency band gaps in many acoustic applications \cite{brule2014experiments, jimenez2017rainbow}.

Exploiting bandgaps at near-zero frequencies allows for precise control of wave fields, with potential applications in managing low-frequency mechanical waves for seismic, and sound control; the zero-frequency bandgap is named for its forbidden zone starting at zero frequency. It was first studied in the context of the quasi-static approximation of the acoustic band \cite{nicorovici1995photonic} and low frequency plasmons \cite{pendry1996extremely} in metallic photonic crystals. The key fundamental aspect of obtaining a zero-frequency bandgap for phonons is to get either zero stiffness over a range of frequencies starting from zero (which is forbidden by the thermodynamic laws) or to have an infinite mass density (forbidden by the standard model). Most studies on zero-frequency bandgaps have sidestepped this by using clamped holes or pillars with Dirichlet boundary conditions. Theoretical studies on pinned plates have examined the existence of zero-frequency stop-bands to shield elastic waves using clamped holes arranged in square arrays with Dirichlet conditions \cite{antonakakis2014moulding}. Another study explored generating a zero-frequency bandgap using clamped pillars embedded in bedrock based on fixed pillars \cite{achaoui2017clamped}. 
More recently, a Bragg gap at nearly zero frequency has been proposed by introducing an additional degree of freedom represented by spin motion, which interacts with the longitudinal motion of a mass in a diatomic mechanical system \cite{oh2018zero}.

In this Letter, we propose a novel paradigm for creating either zero-frequency and zero-wavenumber stop band gaps through optomechanical interactions. Employing highly focused laser beams, formed with aspherical lenses that act as optical tweezers, we effectively trap masses within optical potentials. In such traps, optical forces contribute additional mechanical forces to the masses, akin to supplemental springs attached to each mass. Furthermore, due to the higher propagation velocities of optical waves compared to elastic waves, optical forces can be applied significantly faster than any comparable acoustic counterpart. This allows for the creation of highly responsive, controllable systems in the spirit of space-time metamaterials, where optical forces can be dynamically activated.
This phenomenon of optomechanical trapping results in the emergence of zero-frequency bandgaps when the effective optical stiffness is positive, and zero-wavenumber bandgaps when it is effectively negative. More intriguingly, using this same type of optical actuation, we have been able to mimic negative springs by modulating the optical traps, resulting in a zero-wavenumber bandgap.

\begin{figure}
    \centering
    \includegraphics[width=8cm]{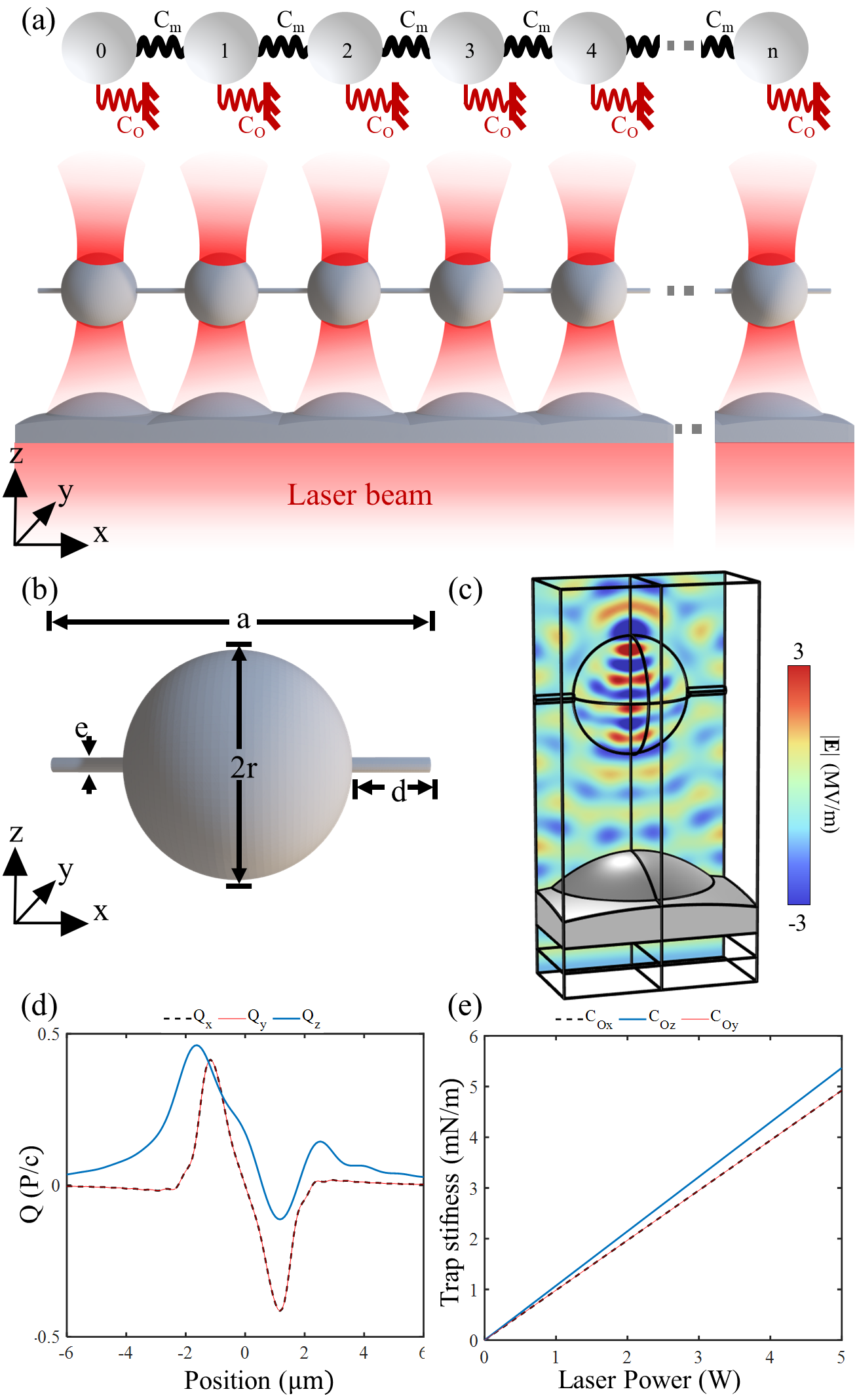}
    \caption{\justifying{Principle of an optomechanically controlled monoatomic mass-spring chain. (a) Illustration of the spherical masses arranged as a monoatomic toy model connected via $C_{m}$ springs and the optical trapping stiffness $C_{O}$ acting on each mass as being attached to the foundation. The lower panel shows the corresponding physically realizable mechanical system composed of spherical masses connected by rods and illuminated via an optical metasurface composed of spherical-like lenses acting individually as traps on the spheres. A linearly polarized optical laser is illuminating the system from the bottom (in the $z$-direction) (b) The geometric representation of the unit cell with a periodicity constant $a =5$ \si{\mu m}, the mass radius $r = 1.5$ \si{\mu m}, the length of the cylinder $d =1$ \si{\mu m} and the diameter $e =0.2$ \si{\mu m}. (c) An illustration of the distribution of the electric field modulus imposed on the spherical mass by the focused laser beam. (d) The trapping efficiencies ($Q_x$, $Q_y$, $Q_z$) for the sphere immersed, are graphed in the transverse $x$, $y$ (black dotted and red lines respectively) and longitudinal $z$ (blue line) directions, as function of the particle displacement from the nominal paraxial focus. (e) The equivalent optical stiffness in the function of the laser power along the 3 principal directions.}}
    \label{Figure 1}
\end{figure}

Our optomechanical model, as shown in \autoref{Figure 1}(a), is constructed as a monoatomic mass-spring chain, and each mass is exposed to optical forces generated by a highly focused laser beam by aspherical lenses with a numerical aperture N.A.$=1.29$ and made up of a material with the refractive index of $n=1.52$: this laser-focused beam, with a wavelength of $1064$\si{nm}, acts as a local optical tweezer. In the trapping condition, the mass acts as if it is attached to another spring, depending on the optical stiffness, which is represented by the equivalent model in the top of \autoref{Figure 1}(a). The geometrical parameters of the mass-spring elementary cell are illustrated in \autoref{Figure 1}(b), in which the periodic lattice constant has a constant of $a = 5$ \si{\mu m} and spherical masses with a radius of $r = 1.5$ \si{\mu m}. For the mechanical calculations reported here we use polymer-like material with material density $\rho = 1100$ \si{\rm kg/m^3}, Young's modulus of $1$ \si{\rm GPa} for masses and $2.7$ \si{\rm MPa} for springs, and Poisson's ratio $\nu = 0.4$.

Firstly, we calculate the optical forces exerted on the masses to verify its practicability and potential experimental verification. We consider a circular polarization for the incident wave with the parameters mentioned earlier and the plane wave becomes focused due to the aspherical lens and creates a Gaussian-like beam. The distribution of the electric field on the mass-spring unit cell is depicted in \autoref{Figure 1}(c). In this figure, we have a full view of the mass and a cross-section in the $xz$-plane for an incident wave propagating upward in the $z$-direction.
Using the outcomes of this simulation, we calculate the trapping efficiencies as a function of the positions $x$, $y$, and $z$ in the mass: $Q_i= {c F_i}/{n_{\rm air} P}$ with $i=$ $x$, $y$, or $z$, $c$ is the light velocity, $P$ is the laser power, $n_{\rm air}$ is the refractive index of air, and $F_i$ is the optical forces (see supplemental materials); the results of the normalized trapping efficiencies are illustrated in \autoref{Figure 1}(d).

At the equilibrium point, the optical force can be linearized as an elastic restoring force with a negative slope, e.g., $F_x = - C_{Ox} x$ for the $x$-direction. Thus, optical forces can be approximated with an effective harmonic potential with spring constants or trap stiffnesses $C_{Ox}$, $C_{Oy}$ and $C_{Oz}$. To calculate the optical trap's stiffness, we simply get the slope of the force-displacement graphs at the equilibrium position, where the force vanishes. In \autoref{Figure 1}(e), we show the calculated stiffness as a function of laser power. 

\begin{figure}[h!]
    \centering
    \includegraphics[width=8.8cm]{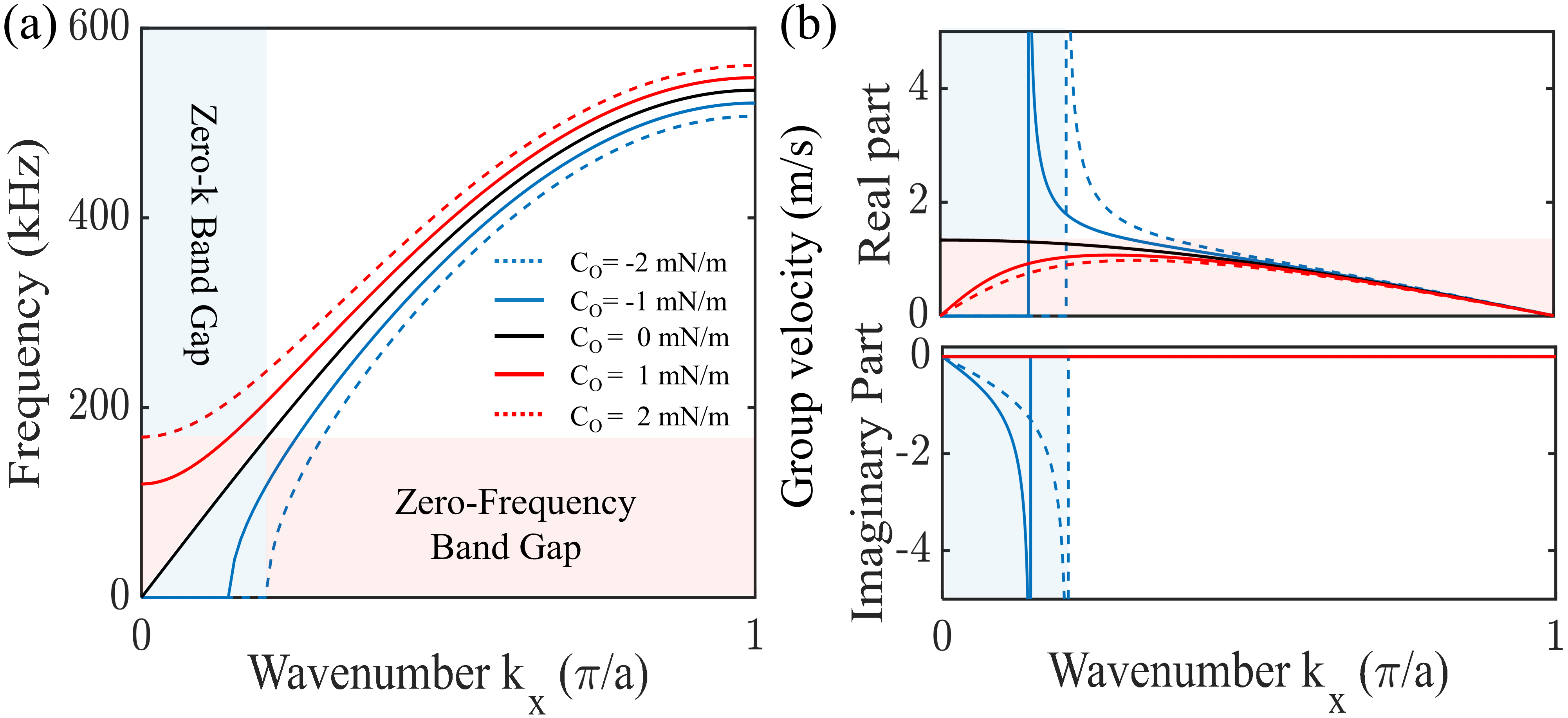}
    \caption{\justifying{ (a) Analytical phononic dispersion curves in the first irreducible Brillouin zone along x-direction as a function of the stiffness of the optical traps $C_O$ on the mass-spring model, with $C_m=0.01$ \si{N/m}, $a=5$ \si{\mu m}, and the Bloch wavenumber $k_x\in [0, \pi/a]$}. (b) Corresponding real and imaginary parts of the group velocity $v_g = \frac{d\omega}{dk}$ are depicted versus wavenumber $k_x$.}
    \label{Figure 2}
\end{figure}

Secondly, we derive the dispersion relation for the mass-spring chain by starting with an analytical model where optical mass trapping is equated to a monoatomic mass-spring system. In this model, each mass is subjected to a highly focused laser that exerts Maxwell stress in the $z$-direction. This setup is analogous to a classical mechanical system where a mass is attached to an additional spring, with the spring's stiffness determined by the optical forces applied by the laser trap. \autoref{Figure 1}(a) illustrates the optical trapping mechanism and its equivalent model for $n$ masses.

To calculate the dispersion diagrams, we analyze the motion of the masses in the system mentioned before. By extracting the forces applied to each mass in the chain, we write the Lagrangian for the mechanical system, which consists of $n$ masses connected by springs with a spring constant $C_m$, and each mass attached to a clamping spring with stiffness $C_O = C_{Ox}$ depending on the optical force applied to that mass. The Lagrangian is expressed as follows:
\begin{equation}
\begin{array}{c c}
    & {L}=\frac{1}{2} \sum_{i=0}^n \left[ m  \dot{u}_i^2\right. \\
    &+\left.C_m\left(\left(u_{i+1}- u_i\right)^2-\left(u_i-u_{i-1}\right)^2\right)-C_o u_i^2\right],
\end{array}
\end{equation}
where $u_i$ is the displacement of mass $m_i$ from its equilibrium position, $\dot{u}_i$ its time derivative, $m$ is the mass of each particle in the chain (assuming all masses are identical), $C_m$ is the spring constant of the springs connecting each mass to its neighbors, $u_{i-1}$ and $u_{i+1}$ are the displacements of the neighboring masses, and $C_O$ is the spring constant describing the optical force.

To solve this system of equations, we need to derive the Lagrangian equation for each degree of freedom, as follows: $\frac{d}{d t}\left(\frac{d L}{d \dot{u_i}}\right)-\frac{\partial L}{\partial u_i}=0$. 
In addition, we assume Bloch waves in the system, as follows:
\begin{equation}\label{eq00002}
    u_i(t) = U_i e^{j (n{k}_x{a} - \omega t)},
\end{equation}
where {$\mathrm{j}^2=-1$}, $U_i$ {($i=1,\cdots$, n)} is the amplitude of $u_i$, ${k}_x$ the component of the wavevector $\textbf{k}$ along $x$, {${a}$ the pitch of the array} and $\omega$ the angular frequency.

Solving these equations (see supplemental materials), we obtain the dispersion relation of the system:

\begin{equation}
\omega=\sqrt{\frac{2}{m} \left(C_m \sin^2 \left(\frac{k_x a}{2}\right)+C_O \right)}.
\end{equation}

In \autoref{Figure 2}, we plot the analytical dispersion of the longitudinal wave of the monatomic mass-spring system, where each mass is connected to an optical stiffness $C_O$. We depict the phononic dispersion along an edge of the first irreducible Brillouin zone ($\Gamma X$), where $\Gamma=(0, 0, 0)$ and $ X=(\pi/a, 0, 0)$, assuming that the cells are repeated periodically along the $x$-direction and finite in the $yz$-plane, making use of the rotational symmetry of the spheres and beams, and aware of dangers of using edges of the Brillouin zone \cite{craster2012dangers}. The first mode, represented in a black line, begins at zero frequency and has an optical stiffness constant set to zero. This shows typical dispersion without a zero-frequency band gap. In other words, the group velocity is not zero; therefore, any wave polarization travels through the mass-spring system. This behavior is typical of a system in which wave propagation is dependent on the material properties, allowing all waves to pass through the medium. However, when an optical stiffness of $1$ \si{mN/m} is added to the mass, as indicated by the red line curve, a zero-frequency band gap appears. This indicates that the group velocity equals zero at these frequencies, forbidding wave propagation over the zero-stop-band gap. As a result, no wave will pass through the system at these frequencies, demonstrating the presence of a zero bandgap caused by the additional optical springs.  This shows that applying forces equivalent to a constant spring stiffness introduces a zero-frequency band gap. Further, when additional forces are applied to the mass, such as in the red dotted line curve where $C_O$ is $2$ \si{mN/m}, the zero-frequency band gap becomes larger, and the longitudinal mode at $k_x=0$ shifts to higher frequencies. This is an analytical demonstration that applying optical forces to the masses results in the formation of a zero-frequency band gap in a monoatomic mass-spring chain.

Another remarkable phenomenon occurs when the stiffness of one of the optical forces becomes negative, specifically reaching values of $-1$ \si{mN/m} and $-2$ \si{mN/m} in the case of the blue dotted line and line curves, respectively. This negative stiffness induces the formation of a zero-wavenumber band gap. In this band gap, the group velocity, which describes the speed at which elastic waves propagate through the mass-spring system, diverges to infinity. This means that within this band gap, the wave packets associated with the optical forces can theoretically travel instantaneously, leading to unique and potentially useful physical properties.

\begin{figure}[h]
    \centering
    \includegraphics[width=8cm]{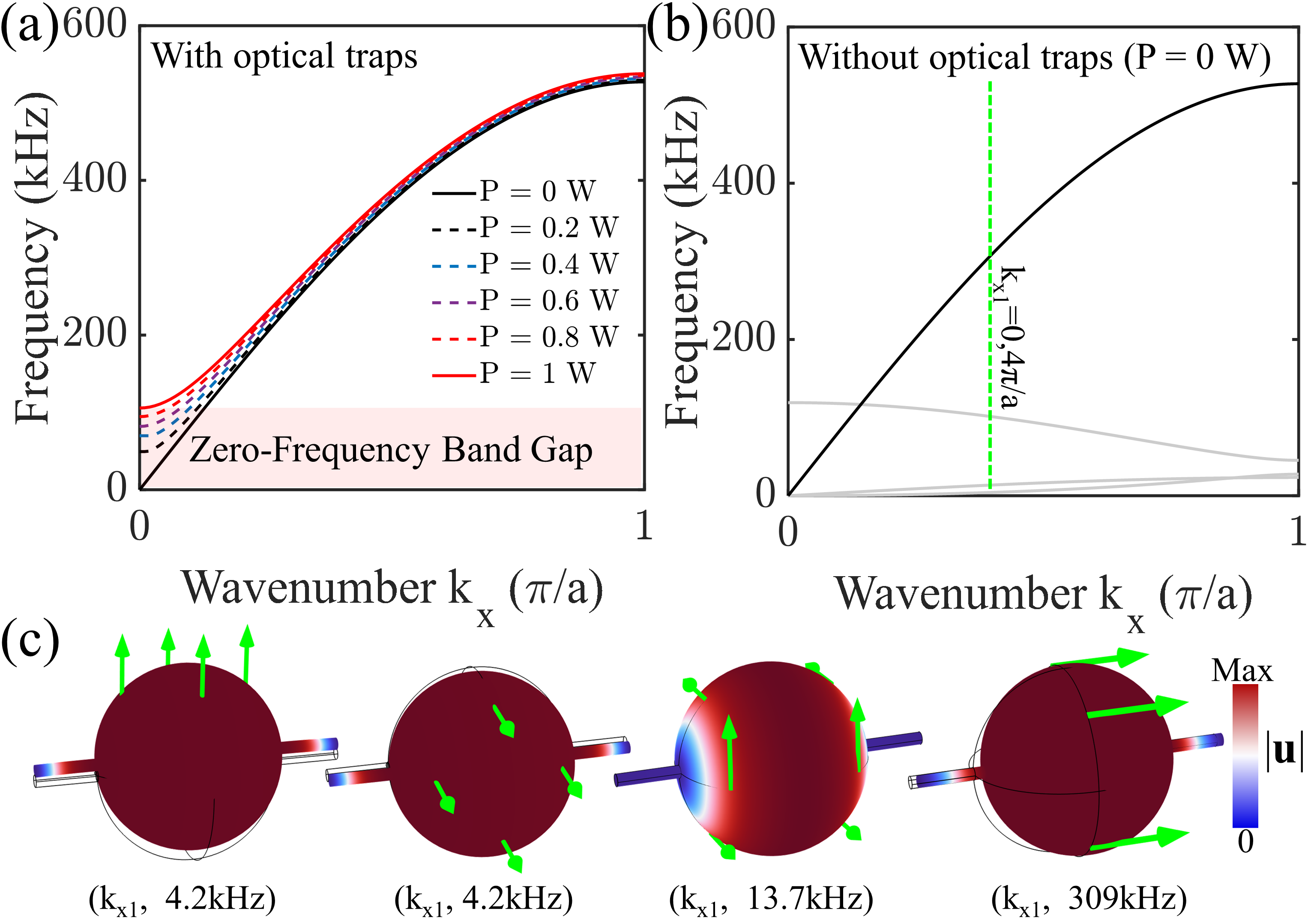}
    \caption{\justifying{Phononic dispersion curves in the first irreducible Brillouin zone $\Gamma \rm X$ for infinite structure. (a) The dispersion curves without optical traps. The longitudinal mode is highlighted in black, while the other modes are shown in gray. (b) The evolution of the longitudinal mode as a function of laser power. The stronger the Laser power P, the wider the zero-frequency stop band. (c) Screenshots of the eigenmodes at $k_{x1}=0.4\pi/a$ display the four fundamental modes: two degenerate transverse modes, a torsional mode, and a longitudinal mode, respectively.}}
    \label{Figure 3}
\end{figure}

After calculating the analytical phononic dispersion in the first Brillouin zone along the $x$-direction, we numerically calculate its counterpart using the finite element method using the elementary cell shown in \autoref{Figure 1}(b), taking into account the optical stiffness calculated for several values of the laser power up to $1$ \si{W}. %


\autoref{Figure 3} depicts the results of eigenvalue calculations. Several major conclusions can be drawn from this figure: at zero frequency, there is no longitudinal mode, indicating that any longitudinally polarized elastic wave cannot propagate at these near-zero frequencies. 

Furthermore, we conduct a sweep calculation of the eigenfrequencies of each mode as a function of the laser power; the results of this investigation are illustrated in \autoref{Figure 3}(a). As the power of the laser increases the upper range of the zero-frequency band gap of each mode shifts to higher frequency;  the frequency range of the zero-gap is controlled by the altering laser power.

Until now we have demonstrated the existence of the zero-frequency and zero-wavenumber band gaps in an infinite array of a monoatomic mass-spring chain. Now, we proceed to calculate the phononic diagrams and demonstrate the existence of the zero-gaps in the case of a finite array composed of $n=40$ elementary cells, which is an experimental approach to demonstrate the existence of the zero-frequency band gap. The schematic illustration of this array is presented in \autoref{Figure 1}(a) with $n=40$. We take into account the optical stiffness applied to each mass by the highly focused laser beam. We use the time domain study to solve the problem, by exciting the finite array with a noise signal, and at each of the $i = 0, 1, \dots, 39$ masses we evaluate the displacement amplitude $|U_{x, y, \ \text{or}\ z}(t)|$ over real-time $t$. Then we conduct the 2D Fast Fourier Transformation from time-real space to frequency-wavenumber, and we obtain $40$ points within the first Brillouin zone $k_x \in [-\pi / a, \pi / a]$. To ensure consistency across frequencies, we normalize the displacement amplitude to a unity power density such that $\sum_{i=0}^{39} \lvert U_i(k_x, \omega) \rvert^2 = 1$. Furthermore, to better understand the zero gap, we calculate the transmission between the fast Fourier transformation of the displacements at the first mass ($i=0$) and the last mass ($i=39$), by using the expression for the transmission, T, as ${\rm T}0=20 \, \log_{10} \left( \frac{|U_{39}(\omega)|}{|U_0(\omega)|} \right)$.
\begin{figure}[!h]
    \centering
    \includegraphics[width=8.7cm]{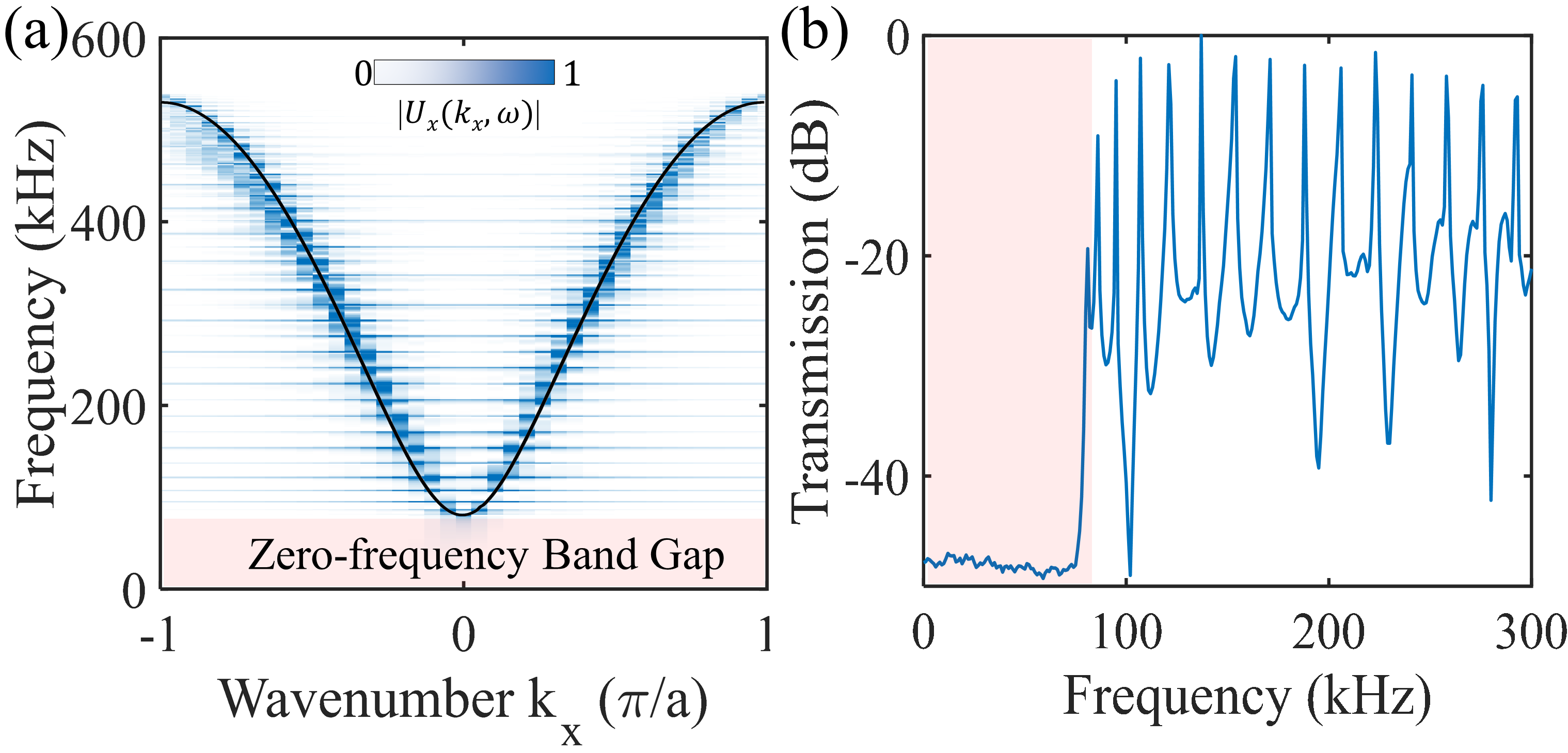}
    \caption{\justifying{Finite mass-spring chain optically trapped in the temporal domain. (a) The calculated phononic dispersion of a finite mass-spring chain under optical traps for $40$ unit cells along the $x$-axis. The modulus $|U_x(k_x, \omega)|$ of the complex response function is plotted on a false-color scale versus the $x$-component $k_x$ of the wave vector and frequency $f=\omega/2\pi$. The black solid line represents the numerical computation of dispersion for an infinite mass-spring chain. (b) The calculated transmission in dB between the modulus of the input $x$-component displacement and the modulus of the output  after $40$ unit cells $x$-component displacement.}}
   \label{Figure 4}
\end{figure}

\autoref{Figure 4} presents the calculated band structures for the infinite array of mass-spring chains within the first Brillouin zone and its calculated transmission over the frequency. The phononic dispersion results presented in \autoref{Figure 4}(a) are displayed on false-color scales, additionally, black solid curves represent band structure calculations for an infinitely periodic array along the $y$-direction.

In the phononic dispersion, we plot three fundamental modes: bending and transverse modes and the longitudinal mode. One can see that all modes have zero frequency band gaps, which is demonstrated by the transmission calculation in \autoref{Figure 4}(b), in which the transmission still equals $-47$ dB along the $0-80$ kHz frequency range. 

\begin{figure}[!h]
    \centering
    \includegraphics[width=5cm]{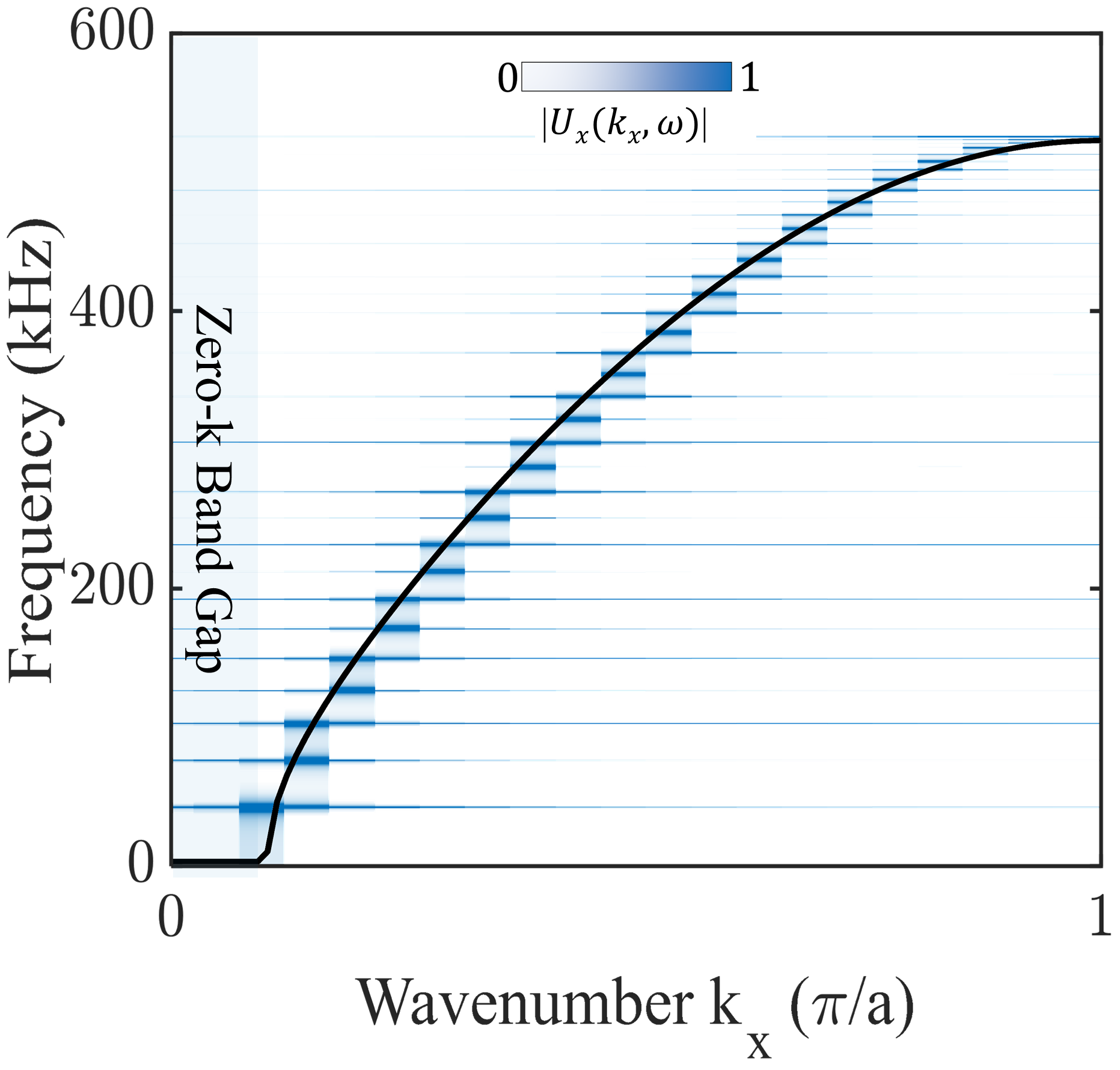}
    \caption{\justifying{Finite mass-spring chain optically driven in the temporal domain. The calculated phononic dispersion of a finite mass-spring chain under optical traps for $40$ unit cells along the $x$-axis. }}
   \label{Figure 5}
\end{figure}
Lastly, we now actively trap and move the previously explained traps in order to control, like in space-time metamaterials, the applied stiffness on the masses. Doing it in this configuration, we can make a counteracting spring leading to a negative stiffness. The corresponding dispersion relation is given in \autoref{Figure 5}. We see a clear signature of a zero wave number region from 0 to finite wave number where no possible mode could be excited to fit a small wavenumber (i.e. large wavelength).


In conclusion, we present an approach that uses optical tweezers to trap and precisely manipulate masses in mass-spring arrays, creating a zero-frequency band gap and also a zero-wavenumber gap. This study introduces an experimental paradigm for generating these band gaps via optomechanical interactions, offering an alternative to conventional methods that use clamped pillars or holes in the zero-frequency case. A challenge for time-varying metamaterials, where the time period of the material modulation is desired to be similar, or less, than that of the wavelength is to modulate a material fast enough; in optics this requires femtosecond modulation pushing the boundaries of experimental laser physics \cite{Tirole2023}. The technique used here in the simple mechanical system, using local optical tweezers, allows effective manipulation of matter on timescales much shorter than the waves in the chain: this opens the way to validating theoretical results for time-modulated metamaterials.

\title{Supplementary Information for: Space-Time Elastic Metamaterials for Zero-frequency and Zero-wavenumber Bandgaps}


\author{Brahim Lemkalli}
\email{lemkallibrahim@gmail.com}
\affiliation{Université de Franche-Comt\'{e}, Institut FEMTO-ST, CNRS UMR 6174, Besan\c{c}on, 25000, France}
\author{Alaa Ali}
\affiliation{Université de Franche-Comt\'{e}, Institut FEMTO-ST, CNRS UMR 6174, Besan\c{c}on, 25000, France}
\author{Qingxiang Ji}
\affiliation{Université de Franche-Comt\'{e}, Institut FEMTO-ST, CNRS UMR 6174, Besan\c{c}on, 25000, France}
\author{Julio Andrés Iglesias Martínez}
\affiliation{Université de Lorraine, Institut Jean Lamour, CNRS UMR 7198, Nancy 54000, France}
\author{Younes Achaoui}
\affiliation{OPTIMEE Laboratory, Department of Physics, Moulay Ismail University, B.P. 11201, Zitoune, Meknes, Morocco}
\author{Sebastien Guenneau}
\affiliation{UMI 2004 Abraham de Moivre-CNRS, Imperial College London, SW7~2AZ, London, UK}
\author{Richard Craster}
\affiliation{UMI 2004 Abraham de Moivre-CNRS, Imperial College London, SW7~2AZ, London, UK}
\author{Muamer Kadic}
\affiliation{Université de Franche-Comt\'{e}, Institut FEMTO-ST, CNRS UMR 6174, Besan\c{c}on, 25000, France}

\maketitle

\section{Supplementary Note 1: Mechanisms of optical trapping}

In this section, we simplify optomechanical interaction by taking into account the orders of magnitude of the different frequencies related to the problem of optical oscillation against mass-spring vibrations. For the optical part, we consider a tweezer trap based on a laser operating at a wavelength of $\lambda = 1064$ \si{nm}, which corresponds to a frequency of $3 \times 10^{14}$. For the acoustical problem, it is limited to frequencies within the \si{MHz} range due to the lattice constant of $5$ \si{\mu m}. As a consequence, we could solve the electromagnetic problem independently and apply the outcome as a constant trapping force in the acoustic problem. The highly focused laser beam used in this work is like Gaussian with circular polarization, and its electric field can be expressed as follows:

\begin{equation}\label{eq01}
\mathbf{E}=
\mathbf{E}_0\frac{w_0}{w(x)} \exp \left[-\frac{y^2+z^2}{w^2(x)}-j q x-j q \frac{y^2+z^2}{2 R(x)}+j \eta(x)\right],
\end{equation}

where $j^2=-1$, $\mathbf{E}_0= E_0 (0,1,j)^T$, $w_0$ is the beam waist which is the radius at the focus, $p_0$ is the position of the focal plane which is zero in our case,  $q$ is the optical wavenumber, $w(x)$ is the radius at specific $x$ position, and it is calculated as:

\begin{equation}\label{w}
w(x)=w_0 \sqrt{1+\left(\frac{x-p_0}{x_0}\right)^2},
\end{equation}

where $x_0=\frac{q w_0^2}{2}$ and $R(x)$ is the radius of curvature of the beam's wavefront, and it is computed as:

\begin{equation}\label{R}
R(x)=\left(x-p_0\right)\left[1+\left(\frac{x_0}{x-p_0}\right)^2\right],
\end{equation}

$\eta(x)$ is the Gouy phase which is defined as a phase advance to the beam around the focal point which results in an increase in the wavelength near the waist; therefore, the phase velocity exceeds the speed of light. The Gouy phase is calculated according to equation \ref{gouy}:

\begin{equation}\label{gouy}
\eta(x)=\tan ^{-1}\left(\frac{x-p_0}{x_0}\right)
\end{equation}

The radiation pressure is then calculated using Maxwell's stress tensor, a parameter depending on the electric and magnetic field components of the total field (incident and scattered).

Generally, the stress is a 2nd-rank symmetric tensor depicted as:

\begin{equation}\label{tensor}
{\mathbf{T}}=\left[\begin{array}{lll}
T_{x x} & T_{x y} & T_{x z} \\
T_{x y} & T_{y y} & T_{y z} \\
T_{x z} & T_{y z} & T_{z z}
\end{array}\right]
\end{equation} 

Each component of this tensor can be computed from the previous field's components:
\begin{equation}
\left\{
\begin{array}{cc}
T_{xx}&= \Re[\epsilon_0 (E_x E^*_x-\frac{1}{2} {E}^2)+\frac{1}{\mu_0}(B_x  B^*_x-\frac{1}{2}{B}^2)]\\
T_{yy}&= \Re[\epsilon_0 (E_y E^*_y-\frac{1}{2} {E}^2)+\frac{1}{\mu_0}(B_y  B^*_y-\frac{1}{2} {B}^2)]\\
T_{zz}&= \Re[\epsilon_0 (E_z E^*_z-\frac{1}{2} {E}^2)+\frac{1}{\mu_0}(B_z B^*_z-\frac{1}{2}{B}^2)] \\
T_{xy}&= \Re[\epsilon_0 (E_x E^*_y)+\frac{1}{\mu_0}(B_x  B^*_y)] \\
T_{xz}&= \Re[\epsilon_0 (E_x E^*_z)+\frac{1}{\mu_0}(B_x  B^*_z)] \\
 T_{yz}&= \Re[\epsilon_0 (E_y E^*_z)+\frac{1}{\mu_0}(B_y B^*_z)] 
\end{array}
\right.
\end{equation}
where ${B}^2:=\mathbf{B}\cdot\mathbf{B}^*$ and ${E}^2:=\mathbf{E}\cdot\mathbf{E}^*$.

\begin{figure}[!h]
    \centering
    \includegraphics[width=0.8\linewidth]{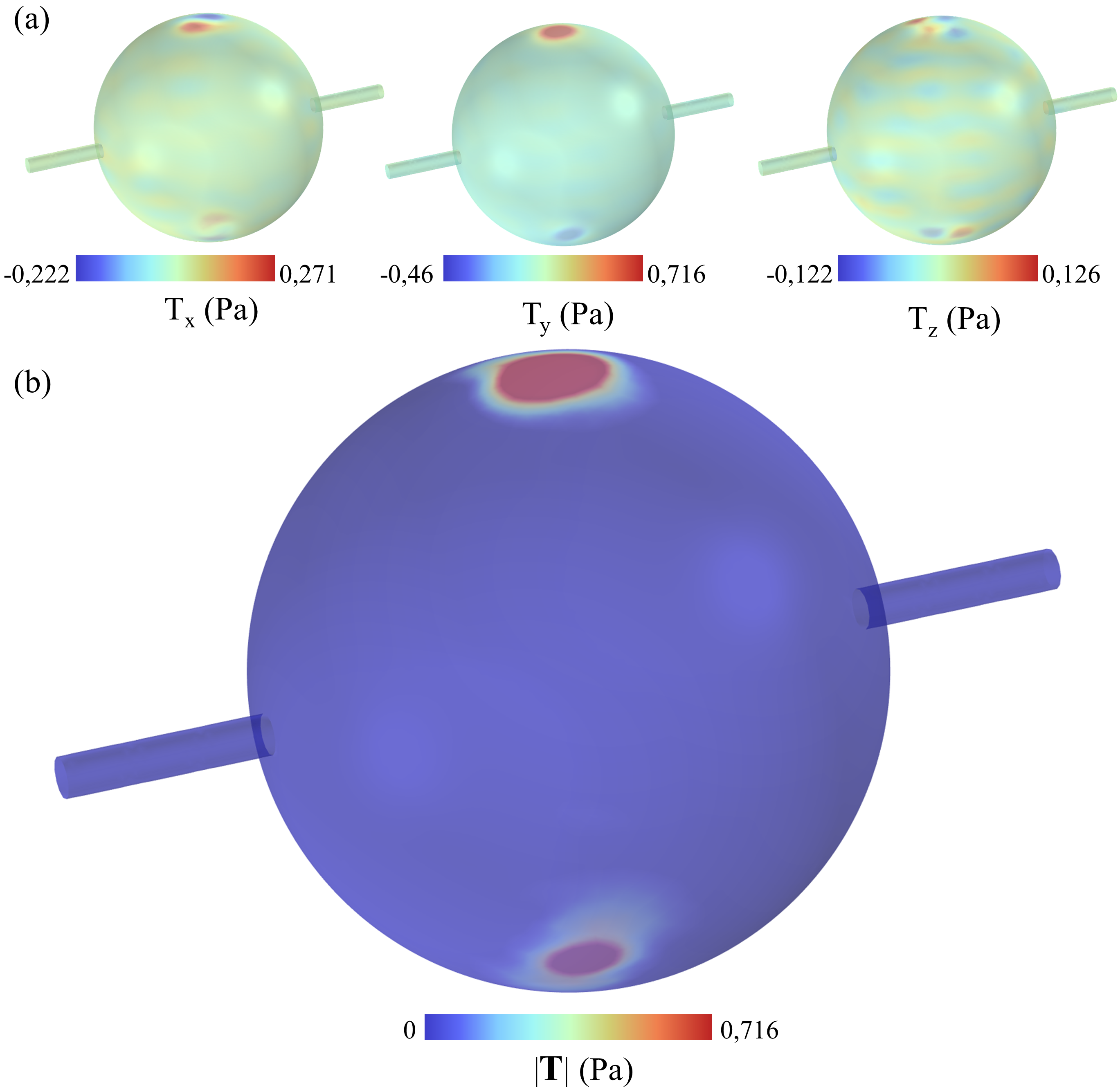}
    \caption{Maxwell stress tensor distribution calculated by using an optical lens that focuses the laser beam on the mass-spring system. (a) The Maxwell stress tensor's $x$, $y$, and $z$-components. (b) The Maxwell Stress Tensor's modulus.}
    \label{Figure MST}
\end{figure}

The total optical force is computed on a particle by integrating the stresses over the external surface using the equation $\mathbf{F}  =  \iint_S \mathbf{T} \cdot d\mathbf{S}$. 

The theoretical description of optical forces depends on the comparison between the spherical mass radius $r$ and the gaussian laser wavelength $\lambda$. Optical trapping can be described in three main regimes based on whether the particle size is smaller than, comparable to, or larger than the wavelength. In our case, with a laser wavelength $\lambda$ of $1$ $\mu$m and a mass radius of approximately $1.5$ $\mu$m, the particle size is significantly larger than the wavelength. Therefore, we use the geometrical optics approximation.

According to ray-optics theory, the maximum restoring radial or axial force occurs when the sphere is displaced from the focal point by a distance equal to its radius. Specifically, for the transverse force along the $r$-direction, the force is given by \cite{malagnino2002measurements, polimeno2018optical}:

\begin{equation}\label{eqq1}
F = C_{Or} r, 
\end{equation}
Where $r$ is $x$ or $y$. Similarly, for the axial force along the $z$-direction, the force is given by:

\begin{equation}\label{eqq2}
F = C_{Oz} r
\end{equation}

Here, $C_{Or}$ and $C_{Oz}$ are the radial and axial stiffness coefficients of the trap, respectively. The main relationship between the trapping force (axial or radial) and the laser power is:

\begin{equation}\label{eqq3}
F = \frac{Q_i  n  P}{c} ,
\end{equation}

where $Q_i$ is a dimensionless parameter that describes the radial and axial efficiency of the trap, representing the fraction of the laser beam's momentum $\frac{nP}{c}$ that is converted into the trapping force, for air the refractive index is equal to $n=1$. Then, by using \ref{eqq1}, \ref{eqq2} and \ref{eqq3} we have the expressions of the optical stiffness as follows:

\begin{equation}
C_{Ox}= \frac{Q_x}{r}\frac{P}{c},
\end{equation}
\begin{equation}
C_{Oy}= \frac{Q_y}{r}\frac{P}{c},
\end{equation}
and
\begin{equation}
C_{Oz} = \frac{Q_z}{r}\frac{P}{c} 
\end{equation}

\section{Supplementary Note 2: Analytical phononic dispersion calculations}
In this section, we give the detail of deriving the analytical dispersion relation for the mass-spring chain where optical mass trapping is equated to a monoatomic mass-spring system. In this model, each mass is subjected to a highly focused laser that exerts Maxwell stress in the y-direction. This setup is analogous to a classical mechanical system where a mass is attached to an additional spring, with the spring's stiffness determined by the optical forces applied by the laser trap.

To calculate the dispersion diagrams, we analyze the motion of the masses in the system mentioned before. By extracting the forces applied to each mass in the chain, we can write the Lagrangian for the mechanical system, which consists of $n$ masses connected by springs with a spring constant $C_m$, and each mass attached to a clamping spring with stiffness $C_O$ depending on the optical force applied to that mass. The Lagrangian is expressed as follows:
\begin{equation}
\begin{array}{c c}
    & {L}=\frac{1}{2} \sum_{i=0}^n \left[ m  \dot{u}_i^2\right. \\
    &+\left.C_m\left(\left(u_{i+1}- u_i\right)^2-\left(u_i-u_{i-1}\right)^2\right)-C_o u_i^2\right],
\end{array}
\end{equation}
where $u_i$ is the displacement of mass $m_i$ from its equilibrium position, $m$ is the mass of each particle in the chain (assuming all masses are identical), $C_m$ is the spring constant of the springs connecting each mass to its neighbors, $u_{i-1}$ and $u_{i+1}$ are the displacements of the neighboring masses, and $C_O$ is the spring constant describing the optical force.

To solve this system of equations, we need to derive the Lagrangian equation for each degree of freedom, as follows: $\frac{d}{d t}\left(\frac{d L}{d \dot{u_i}}\right)-\frac{\partial L}{\partial u_i}=0$. Then, for each mass $m_i$ in the chain, the equation of motion can be expressed as:

\begin{equation}\label{eq00001}
\left\{
\begin{array}{c}
    m \ddot{u}_0 = C_m (u_{1} + u_{-1}-2u_{0}) -C_O u_0\\
    m \ddot{u}_1 = C_m (u_{2} + u_{0}-2u_{1}) -C_O u_1\\
    \vdots\\
    m \ddot{u}_n = C_m (u_{n+1} + u_{n-1}-2u_{n}) -C_O u_n
\end{array}
\right..
\end{equation}

The monoatomic chain is periodic with the lattice constant $a$, representing the distance between two nearest masses. The solutions of the system of equations \eqref{eq00001} are written as Bloch waves, as follows:

\begin{equation}\label{eq00002}
    u_i(t) = U_i e^{j(n{k}_xa - \omega t)},
\end{equation}
where $U_i$ is the amplitude, $k_x$ is the component of the wavevector $\textbf{k}$ along $x$, and $\omega$ is the angular frequency. The differential equations, along with these boundary conditions, fully describe the dynamics of the system of masses connected by springs in the infinite chain. 

Solving these equations would provide the analytical solution for the displacement $u_i$ of each mass in the chain as a function of wavenumber. Replacing the plane wave solution \eqref{eq00002} into the equation of motion \eqref{eq00001} results in the following system of equations:

\begin{equation}\label{eq12}
\left\{
\begin{array}{c}
    -\omega^2 m U_0 = C_m (U_{1}e^{jk_xa} + U_{-1}e^{-jk_xa}-2U_{0}) -C_O U_0\\
    -\omega^2 m U_1 = C_m (U_{2}e^{jk_xa} + U_{0}e^{-jk_xa}-2U_{1}) -C_O U_1\\
    \vdots\\
    -\omega^2 m U_n = C_m (U_{n+1}e^{jk_xa} + U_{n-1}e^{-jk_xa}-2U_{n}) -C_O U_n\\
\end{array}
\right.
\end{equation}
which is nothing but an eigenvalue problem $C(k_x)\textbf{U}=\omega^2 M\textbf{U}$, where $\textbf{U}$ is the amplitude vector, $M$ is the mass matrix ($M_{ij, i=j}=m $), and $C(k_x)$ is a stiffness matrix denoted as follows:
\begin{equation}
\setlength\arraycolsep{0.5pt}
    C(k_x)=\left[
\begin{array}{ccccc}
     2C_m+C_O& -C_m e^{jk_xa}&0 &\hdots&0  \\
    -C_m e^{-jk_xa}& 2C_m+C_O&-Ce^{jk_xa} &\hdots&0\\
    0& -C_m e^{-ik_xa} &2C_m+C_O &\hdots&0\\
    \vdots&\vdots&\vdots&\ddots&\vdots\\
    0&0&0& \hdots& 2C_m+C_O
\end{array}
\right]
\end{equation}

To calculate the dispersion curves arising from (\ref{eq12}), we look for roots of the determinant
$det(C(k_x)-\omega^2 M)=0$.

To streamline the calculation of the system of equations (\ref{eq12}), we selected two nearest neighbor masses $i$ and $i+1$, respectively. Then, we calculate the dispersion of the longitudinal mode by using the expression derived from (\ref{eq12}) as a function of the optical stiffness constant that describes the optical forces applied to both masses. The results of this calculation are illustrated in the Figure 2, where we vary the optical stiffness $C_O$.

\begin{equation}
\omega=\sqrt{\frac{2}{M} \left(C_m \sin^2 \left(\frac{k_x a}{2}\right)+C_O \right)}.
\end{equation}

\section{Supplementary Note 3: Infinite Numerical Phononic dispersion Calculation methods}

In this section, we use the commercial software COMSOL multiphysics to numerically solve the eigenvalue equation derived from Cauchy elasticity, using a similar system to the one studied in the analytical model, by considering an infinite monoatomic mass-spring chain composed of spherical masses with a radius $r$ and $n$ springs attached to these masses with stiffness $C_m$. For enhancing optical trapping in finite element simulations, we use spring foundation boundaries to apply the calculated optical forces to each mass. Then, we solve the elasticity equation starting with the equilibrium equations.

\begin{equation}
\nabla \cdot \boldsymbol{\sigma} + \mathbf{b} = \mathbf{0}
\end{equation}

Combined with the Caushy elastic equation (Generalized Hooke's Law):
\begin{equation}
\boldsymbol{\sigma} = \mathbf{C} : \boldsymbol{\varepsilon}.
\end{equation}
where the strain-displacement relation:
\begin{equation}\label{stress}
\boldsymbol{\varepsilon} = \frac{1}{2} \left( \nabla \mathbf{u} + (\nabla \mathbf{u})^T \right).
\end{equation}

where $\boldsymbol{\sigma}$ is the stress tensor, $\mathbf{b}$ is the body force vector, $\mathbf{C}$ is the elasticity tensor, $\boldsymbol{\varepsilon}$ is the strain tensor, and $\mathbf{u}$ is the displacement vector.

By substituting the equation (\ref{stress}) into Cauchy elasticity for an isotropic homogeneous linear material, we get the equation for the eigenvalue problems, as follows: 

\begin{equation}
    -\rho \omega^2 \mathbf{u}=\frac{E}{2(1 + \nu)(1 - 2\nu)}  \nabla \nabla \cdot \mathbf{u}
+ \frac{E}{2(1 + \nu)} \nabla^2 \mathbf{u}
\end{equation}

Where $E$ is the Young's modulus, $\nu$ is the Poisson's ratio, $\rho$ is the density, $\omega$ is the angular frequency, and $\textbf{u}$ is the displacement vector.

For the optical forces, we use a spring foundation condition by applying the calculated optical stiffnesses at the center of mass, represented by $C_{Ox}$, $C_{Oy}$, and $C_{Oz}$. We then calculate the phononic dispersion relations relation in the first irreducible Brillouin zone ($\Gamma X$), where $\Gamma=(0, 0, 0)$ and $ X=(\pi/a, 0, 0)$, assuming that the cells are repeated periodically along the $x$-direction presenting beam symmetry. We apply the Floquet-Bloch boundary conditions along the $x$-direction on the unit cell, as follows: 

\begin{equation}
\textbf{u}(x+a,y,z)=\textbf{u}(x,y,z)e^{jk_xa},
\end{equation}
where ${\bf k}=(k_x, 0, 0)$ is the reduced Bloch wavevector and $\textbf{u}(x,y,z)$ is the displacement vector.

\section*{References}
\providecommand{\noopsort}[1]{}\providecommand{\singleletter}[1]{#1}%

\end{document}